\documentstyle[aps,prl,twocolumn,epsfig,floats]{revtex}

\tighten

\newcommand{\ii}{{\rm i}}
\newcommand{\de}{{\rm\,d}}
\newcommand{\e}{{\rm e}}

\newcommand{\im}{\,{\rm Im}\,}
\newcommand{\tr}{\mbox{Tr}\,}
                 
\newcommand{\st}{{\scriptscriptstyle T}}
\newcommand{\sa}{{\scriptscriptstyle A}}

\newcommand{\g}{\gamma}
\newcommand{\sig}{\sigma}
\newcommand{\eps}{\epsilon}
\newcommand{\nn}{\nonumber}

\newcommand{\pslash}{\rlap{/} p}
\newcommand{\kslash}{\rlap{/} k}
\newcommand{\lslash}{\rlap{/} l}
\newcommand{\derslash}{\rlap{/} \partial}
\newcommand{\Aslash}{\rlap{/} A}
\newcommand{\Vslash}{\rlap{/} V}

\begin{document}
 
\twocolumn[\hsize\textwidth\columnwidth\hsize\csname
@twocolumnfalse\endcsname

\title{
\begin{flushright}
\begin{minipage}{4 cm}
\small
VUTH 02-02\\
\end{minipage}
\end{flushright}
Estimate of the Collins fragmentation function in a chiral invariant approach}

\author{A.~Bacchetta$^1$, R.~Kundu$^2$, A.~Metz$^1$, P.J.~Mulders$^1$\\[2mm]}

\address{$^1$ Division of Physics and Astronomy, Faculty of Science, Free
University  \\
De Boelelaan 1081, NL-1081 HV Amsterdam, the Netherlands\\[2mm]
$^2$ Department of Physics, RKMVC College\\
Rahara, North 24 Paraganas, India\\[2mm]
}

\date{January 11, 2002}

\maketitle

\begin{abstract}
We predict the features of the Collins function, which describes the 
fragmentation of a transversely polarized quark into an unpolarized 
hadron, by modeling the fragmentation into pions at a low energy scale. 
We use the chiral invariant approach of Manohar and Georgi, where 
constituent quarks and Goldstone bosons are considered as effective 
degrees of freedom in the non-perturbative regime of QCD.
To test the approach we calculate the unpolarized fragmentation function and 
the transverse momentum distribution of a produced hadron, both of which are 
described reasonably well.
In the case of semi-inclusive deep-inelastic scattering, our estimate of 
the Collins function in connection with the transversity distribution gives 
rise to a transverse single spin asymmetry of the order of 10\%, 
supporting the idea of measuring the transversity distribution of the 
nucleon in this way.
In the case of $e^+ e^-$ annihilation into two hadrons, our model predicts a 
Collins azimuthal asymmetry of about 5\%.

\end{abstract}
\pacs{13.60.Le,13.87.Fh,12.39.Fe}

]

\section{Introduction}
\label{s:one}

The influence of transverse spin and transverse momentum on fragmentation 
processes is at present a largely unexplored
subject.
The Collins fragmentation function~\cite{Collins:1993kk}, 
correlating the transverse spin of the fragmenting quark to the transverse 
momentum of the produced hadron, could give us the first chance to study this
effect. 
Moreover, being chiral-odd, the Collins function 
can be connected to the transversity distribution function, which is
chiral-odd as well, and thus can
allow the measurement of this otherwise elusive property of the nucleon, 
which carries valuable information about the
dynamics of confined quarks.
Beside being chiral-odd, the Collins function is 
also time-reversal odd (T-odd).

In spite of the apparent difficulty in modeling T-odd effects,
in a recent paper~\cite{Bacchetta:2001di} we have
shown that a non-vanishing Collins function can be obtained
through a consistent one-loop calculation, in a description where massive
constituent quarks and pions are the only 
effective degrees of freedom and interact via a simple pseudoscalar
coupling. 

In our previous work~\cite{Bacchetta:2001di} little care 
has been devoted to the phenomenology of the Collins function.
In contrast, our interest here lies in obtaining a reasonable estimate of 
this function and the observable effects induced by it.
At present, only one attempt to theoretically estimate the Collins function
for pions exists~\cite{Artru:1997bh}, 
and little phenomenological information is available from experiments.
The HERMES collaboration reported the first measurements of single spin 
asymmetries in semi-inclusive DIS~\cite{Airapetian:2000tv,Airapetian:2001eg},
giving an indication of a possibly non-zero Collins function. The
Collins function has also been invoked to explain large azimuthal
asymmetries in 
$p p^{\uparrow} \rightarrow \pi X$~\cite{Anselmino:1999pw,Boglione:1999dq}. 
In this case, however, 
the extraction of the function is plagued by large uncertainties, due to the
possible presence of hadronic effects both in the initial and final state, 
and hence does not allow any conclusive statement yet. 
Recently, a phenomenological estimate of the
Collins function has been proposed~\cite{Efremov:2001cz}, 
combining results from the DELPHI, SMC and HERMES experiments.
However, in spite of all the efforts to pin down the Collins function, the
knowledge we have at present is still unsufficient.

In this work we calculate the Collins 
function for pions in a chiral invariant approach at a low energy scale.
We use the model of Manohar and Georgi~\cite{Manohar:1984md},
which incorporates chiral symmetry and its spontaneous breaking, two
important aspects of QCD at low energies.
The spontaneous breaking of chiral symmetry leads to the existence
of (almost massless) Goldstone bosons, which are included as effective degrees 
of freedom in the model.
Quarks appear as further degrees of freedom as well. However, 
in contrast to the current quarks of the QCD Lagrangian, the model uses
massive constituent quarks -- a concept
which has been proven very succesful in many phenomenological models
at hadronic scales.
With the exception of Ref.~\cite{Ji:1993qx}, the implications of a chiral 
invariant interaction for fragmentation functions into Goldstone bosons 
at low energy scales remains essentially unexplored.
To investigate the Collins function for vector mesons like the $\rho$
\cite{Czyzewski:1996ih} is beyond the reach of the approach.

Although the applicability of the Manohar-Georgi model
 is restricted to energies below the scale of chiral symmetry 
breaking $\Lambda_\chi \approx 1\, \rm{GeV}$, this is sufficient to calculate
soft objects.
In this kinematical regime, the chiral power counting allows setting up a 
consistent perturbation theory~\cite{Weinberg:1979kz}.
The relevant expansion parameter is given by $l/\Lambda_\chi$, where $l$ is a 
generic external momentum of a particle participating in the fragmentation.
To guarantee the convergence of the perturbation theory, we restrict the
maximum virtuality $\mu^2$ of the decaying quark to a soft value.
We mostly consider the case $\mu^2 =  1 \, \rm{GeV}^2$.

The outline of the paper is as follows: We first give the details of our 
model and present the analytical results of our calculation. 
Next, we discuss our results and compare them with known observables,
indicating the choice of the parameters of our model. Then, 
we present the features
of our prediction for the Collins function and its moments. Finally, using the
outcome of our model, we estimate the leading order asymmetries containing 
the Collins function in semi-inclusive DIS and in $e^+e^-$
annihilation into two hadrons.

\section{Calculation of the Collins function}
\label{s:two}

Considering the fragmentation process of a quark into a pion, 
$q^{\ast}(k) \to \pi(p) X$, we use the expressions of
 the unpolarized fragmentation function 
$D_1$ and the Collins function $H_1^\perp$ in terms of light-cone correlators,
depending on the longitudinal 
momentum fraction $z$ of the pion and the transverse momentum $k_{\st}$ of the 
quark. The definitions read~\cite{Levelt:1994np,Mulders:1996dh}
\footnote{Note that this definition of $H_1^{\perp}$ slightly differs from the
original one given by Collins \cite{Collins:1993kk}.}
\begin{eqnarray} 
&&D_1(z,z^2 k^2_{\st})= \left. 
 \frac{1}{4z} \int \de k^+ \;
              \tr[ \Delta (k,p) \g^-] \right|_{k^-=\frac{p^-}{z}} \,,
\label{e:d1} \\
&&\frac{\eps_{\st}^{ij} k_{\st\,j}}{m_{\pi}}
	      \, H_1^{\perp}(z,z^2 k^2_{\st}) \nn\\
&&\quad\quad\quad\quad\mbox{}= \left. \frac{1}{4z} \int \de k^+ \;
              \tr[ \Delta (k,p) \ii \sig^{i-}\g_5]
	\right|_{k^-=\frac{p^-}{z}} \,,  
\label{e:col1}
\end{eqnarray}
with $m_{\pi}$ denoting the pion mass and $\eps_{\st}^{ij} \equiv \eps^{ij-+}$
(we specify the plus and minus lightcone components of a generic 4-vector 
$a^{\mu}$ 
according to $a^{\pm} \equiv (a^0 \pm a^3)/\sqrt{2}$).
The correlation function $\Delta(k,p)$ in Eqs.~(\ref{e:d1}, \ref{e:col1}), omitting 
gauge links, takes the form
\begin{eqnarray}  
\Delta(k,p)=\sum_X \, \int &&
        \frac{\de^{4}\xi}{(2\pi)^{4}}\; \e^{+\ii k \cdot \xi}
       \langle 0|
\,\psi(\xi)|\pi, X\rangle 
\nn\\
&&\mbox{}\times
\langle \pi, X|
             \bar{\psi}(0)|0\rangle \,.    
\label{e:delta}
\end{eqnarray} 

We now use the chiral invariant model of Manohar and Georgi~\cite{Manohar:1984md}
to calculate the matrix elements in the correlation function.
Neglecting the part that describes free Goldstone bosons, 
the Lagrangian of the model reads
\begin{equation}
{\cal L} =  \bar{\psi} \, ( \ii \derslash + \Vslash - m 
 + g_{\sa} \Aslash \g_5 ) \, \psi \,.
\label{e:lagrangian}
\end{equation}
In Eq.~(\ref{e:lagrangian}) the pion field enters through the vector and 
axial vector combinations 
\begin{equation} 
V^{\mu}  =  \frac{\ii}{2} \, [u^{\dagger} , \partial^{\mu} u] \,,\qquad
A^{\mu}  =  \frac{\ii}{2} \, \{u^{\dagger} , \partial^{\mu} u\} \,,
\end{equation} 
with $u = \textrm{exp}(\ii \, \vec{\tau} \! \cdot \! \vec{\pi} / 2 F_\pi)$,
where the $\tau_i$ are the generators of the SU(2) flavour group and 
$F_\pi = 93 \, \textrm{MeV}$ represents the pion decay constant.
In absence of resonances, the pion decay constant determines the
scale of chiral symmetry breaking via $\Lambda_\chi = 4 \pi F_\pi$.
The quark mass $m$ and the axial coupling constant $g_{\sa}$ are free 
parameters of the model that are not constrained by chiral symmetry.
The values of these parameters will be specified in Sec.~\ref{s:three}.
Though we limit ourselves here to the SU(2) flavour sector of the model, 
the extension to strange quarks 
is straightforward, allowing in 
particular the calculation of kaon fragmentation functions.
For convenience we write down explicitly those terms of the interaction 
part of the Lagrangian (\ref{e:lagrangian}) that are relevant for our 
calculation.
To be specific we need both the interaction of a single pion with a 
quark and the two-pion contact interaction, which can be easily obtained by 
expanding the non-linear representation $u$ in terms of the pion field:
\begin{eqnarray}
{\cal L}_{\pi qq} & = & - \frac{g_\sa}{2 F_{\pi}} \bar{\psi} \gamma_{\mu} 
 \gamma_5 \vec{\tau} \cdot \partial^{\mu} \vec{\pi} \psi \,,
 \label{e:lpi} \\
{\cal L}_{\pi \pi qq} & = & - \frac{1}{4 F_{\pi}^2} \bar{\psi} \gamma_{\mu} 
 \vec{\tau} \cdot ( \vec{\pi} \times \partial^{\mu} \vec{\pi} ) \psi \,.
 \label{e:lpipi}
\end{eqnarray}
Performing the numerical calculation of the Collins function,
 it turns out that 
the contact interaction (\ref{e:lpipi}),
which is a direct consequence of chiral symmetry, plays a dominant role.

\begin{figure}
        \centering
        \epsfig{figure=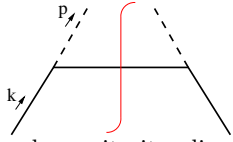,width=3cm}
        \caption{Lowest-order unitarity diagram describing the fragmentation 
		of a quark into a pion.} 
	\label{f:born}
        \end{figure}

At tree level, the fragmentation of a quark is modeled through the process
$q^{\ast} \to \pi q$, where Fig.~\ref{f:born} represents the corresponding 
unitarity diagram.
Using the Lagrangian in Eq.~(\ref{e:lpi}), the correlation function 
at lowest order reads
\begin{eqnarray} 
\Delta_{(0)}(k,p) & = &
 - \frac{g_\sa^2}{4F_\pi^2} \frac{1}{(2\pi)^4}\,
 \frac{(\kslash + m)}{k^2 - m^2} \, \g_5 \, \pslash \, (\kslash - \pslash +m)
 \nn \\
&& \mbox{} \times \pslash \, \g_5\frac{(\kslash + m)}{k^2 - m^2}\, 
 2\pi\,\delta((k-p)^2 -m^2) \,. 
\end{eqnarray} 
This correlation function allows to compute the unpolarized fragmentation 
function $D_1$ by means of Eq.~(\ref{e:d1}), leading to
\begin{eqnarray}
 D_1(z,z^2 k^2_\st) & = & 
 \frac{1}{z} \frac{g_\sa^2}{4 F_\pi^2}
 \frac{1}{16\pi^3} \label{e:d1tree}\\
& & \mbox{} \times \bigg( 1 - 4\frac{1-z}{z^2}
 \frac{m^2 m_\pi^2}{[k_\st^2 +m^2 +\frac {1-z}{z^2} m_\pi^2 ]^2} \bigg). \nn
\end{eqnarray}
Note that the expression in (\ref{e:d1tree}) is only weakly dependent on the 
transverse momentum of the quark.
In fact, $D_1$ is constant as a function of $k_\st$, if $m_\pi = 0$ and (or) $m = 0$. 
Because our approach is limited to a soft regime, we will 
impose an upper cutoff on the $k_\st$ integration, as will be discussed in 
more detail in Sec.~\ref{s:three}.
This in turn leads to a finite $D_1(z)$ after integration over the 
transverse momentum.

The SU(2) flavour structure of our approach implies the relations
\begin{eqnarray} 
D_{1}^{u\rightarrow \pi^0} = D_{1}^{\bar{u}\rightarrow \pi^0} 
 = D_{1}^{d\rightarrow \pi^0} = D_{1}^{\bar{d}\rightarrow \pi^0} & = & D_1 \,,
 \label{e:flavour1} \\
D_{1}^{u\rightarrow \pi^+} = D_{1}^{\bar{d}\rightarrow \pi^+} 
 = D_{1}^{\bar{u}\rightarrow \pi^-} = D_{1}^{d\rightarrow \pi^-} & = & 2 \, D_1 \,,
 \label{e:flavour2}
\end{eqnarray} 
where $D_1$ is the result given in Eq.~(\ref{e:d1tree}).
In the case of unfavoured fragmentation processes $D_1$ vanishes at tree 
level, but will be non-zero as soon as one-loop corrections are included. 
According to the chiral power counting, one-loop contributions to $D_1$ are 
suppressed by a factor $l^2/\Lambda_\chi^2$ compared to the tree level result.
The maximum momentum up to which the chiral perturbation expansion 
converges numerically can only be determined by an explicit calculation of the 
one-loop corrections.

\begin{figure}
	\centering
	\epsfig{figure=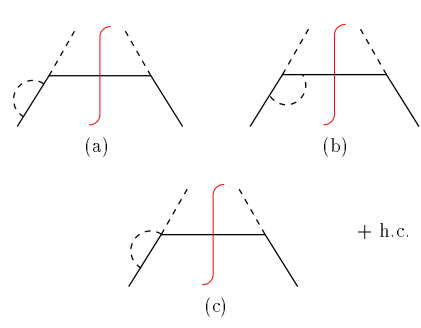,width=8.2cm}
        \caption{One-loop corrections to the fragmentation of a quark
		 into a pion contributing to the Collins function.
                 The hermitian conjugate diagrams (h.c.) are not shown
                 explicitly.}
        \label{f:1loop}
        \end{figure}

\begin{figure}
	\centering
	\epsfig{figure=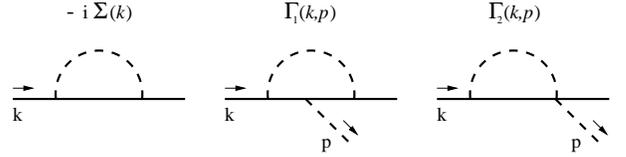,width=8cm}\\
        \caption{One-loop self-energy, and vertex corrections.}
        \label{f:sigmagammadelta}
        \end{figure}

Like in the case of a pseudoscalar quark-pion 
coupling~\cite{Bacchetta:2001di}, the Collins function $H_1^\perp$ turns out to 
be zero in Born approximation.
To obtain a non-zero result, we have to resort to the one-loop level.
In Fig.~\ref{f:1loop} 
the corresponding diagrams are shown, where we have displayed only those 
graphs that contribute to the Collins function.
The explicit calculation of $H_1^\perp$ is similar to our previous 
work~\cite{Bacchetta:2001di}.
The relevant ingredients of the calculation are the self-energy and the vertex 
correction diagrams.
These ingredients are sketched in Fig.~\ref{f:sigmagammadelta} and can 
be expressed analytically as  
\begin{eqnarray}
&& - \ii \Sigma (k) = \frac{g_\sa^2}{4 F_\pi^2}
 \int \frac{\de^4 l}{(2 \pi)^4} \, 
 \frac{\lslash \, (\kslash - \lslash - m) \, \lslash}
 {[(k-l)^2 - m^2]\,[l^2 -m_{\pi}^2]} \,, 
 \\
&& \Gamma_1 (k,p) = - \ii \frac{g_\sa^3}{8 F_\pi^3} \g_5 
 \int \frac{\de^4 l}{(2\pi)^4}
 \nn \\ 
&&\quad \mbox{} \times \frac{\lslash \, (\kslash - \pslash - \lslash + m)}
 {[(k - p - l)^2 - m^2]} \frac{\pslash \, (\kslash - \lslash - m) \, \lslash}
 {[(k - l)^2 - m^2][l^2 - m_{\pi}^2]} \,,
 \\
&& \Gamma_2 (k,p) =  - \ii \frac{g_\sa}{8 F_\pi^3} \g_5 
 \int \frac{\de^4 l}{(2\pi)^4}
 \nn\\ 
&&\quad \mbox{} \times \frac{(\lslash+\pslash)\,(\lslash-\kslash+m)\,\lslash}
{[(k-l)^2 - m^2)] [l^2-m_\pi^2]} \,,
\end{eqnarray} 
where flavour factors have been suppressed.
For later purpose, we give here the most general parametrization
of the functions $\Sigma$, $\Gamma_1$ and $\Gamma_2$,
\begin{eqnarray} 
\Sigma (k) & = & A\,\kslash + B\, m \,, 
 \vphantom{\frac{1}{1}} 
 \label{e:sig} \\
\Gamma_1 (k,p) & = & \frac{g_\sa}{2 F_\pi} \g_5 
 \Big( C_1 + D_1 \,\pslash + E_1 \,\kslash + F_1\,\pslash \, \kslash \Big) \,,
 \label{e:gamma1} \\
\Gamma_2 (k,p) & = & \frac{g_\sa}{2 F_\pi} \g_5
 \Big( C_2 + D_2 \,\pslash + E_2 \,\kslash + F_2\,\pslash \, \kslash \Big) \,.
 \label{e:gamma2}
\end{eqnarray}  
The real parts of the functions $A$, $B$, $C_1$, $D_1$ etc. could be 
UV-divergent and require in principle a proper renormalization.
Here, we don't need to deal with the question of renormalization at all, 
since only the imaginary parts of the loop diagrams are important 
when calculating the Collins function~\cite{Bacchetta:2001di}.

Taking now flavour factors properly into account, the contributions to the 
correlation function generated by the diagrams (a), (b) and (c) in 
Fig.~\ref{f:1loop} are given by
\begin{eqnarray} 
&& \Delta_{(1)}^{(a)}(k,p) = 
 - 3 \frac{g_\sa^2}{4 F_\pi^2} \frac{1}{(2 \pi)^4} \, 
 \frac{(\kslash + m)}{k^2 -m^2} \, \g_5 \,\pslash\, (\kslash - \pslash + m) \, 
 \nn \\
&& \qquad \mbox{} \times \pslash \, \g_5 \frac{(\kslash + m)}{k^2 -m^2} \, 
 \Sigma(k) \, \frac{(\kslash + m)}{k^2 -m^2} \, 
 \nn \\
& & \qquad \mbox{} \times 2 \pi \, \delta ((k-p)^2 -m^2) \,, 
 \vphantom{\frac{1}{1}} 
 \\
&& \Delta_{(1)}^{(b)}(k,p) =  
 \frac{g_\sa}{2F_\pi^2} \frac{1}{(2\pi)^4}\,
 \frac{(\kslash + m)}{k^2 - m^2} \, \g_5 \, \pslash \, (\kslash - \pslash +m)
\nn \\
&& \qquad \mbox{} \times \Gamma_1(k,p) \frac{(\kslash + m)}{k^2 - m^2}\, 
 2 \pi \, \delta ((k-p)^2 -m^2) \,,
 \\ 
&& \Delta_{(1)}^{(c)}(k,p) =
 - 2 \frac{g_\sa}{2F_\pi^2} \frac{1}{(2\pi)^4}\,
 \frac{(\kslash + m)}{k^2 - m^2} \, \g_5 \, \pslash \, (\kslash - \pslash +m)
\nn \\
&& \qquad \mbox{} \times \Gamma_2(k,p) \frac{(\kslash + m)}{k^2 - m^2}\, 
 2 \pi \, \delta ((k-p)^2 -m^2) \,.
\end{eqnarray} 
The correlation functions of the hermitian conjugate diagrams follow from the
hermiticity condition
$\Delta_{(1)}^{h.c.}(k,p)=\gamma^0\Delta_{(1)}^{\dagger}(k,p)\gamma^0$.

Summing the contributions of all diagrams and inserting the resulting
correlation function in Eq.~(\ref{e:col1}), we eventually obtain the result
\begin{eqnarray} 
&& H_1^{\perp}(z,z^2 k^2_{\st}) = 
 \frac{g_\sa^2}{32 \pi^3 F_\pi^2} \frac{m_\pi}{1-z} \frac{1}{k^2 - m^2} 
 \nn \\
&& \quad \mbox{} \times \bigg( - 3m \im \big( A+B \big)
 \nn \\
& & \quad - \im \big( C_1 \! - m E_1 +(k^2 \! - m^2)F_1 \big) 
 \vphantom{\frac{1}{1}}
\label{e:h1}
 \\
&& \quad + 2 \im \big( C_2 \! - m E_2 +(k^2 \! - m^2)F_2 \big) \! \bigg) \!
 \bigg|_{k^2 = \frac{z}{1-z}k_{\st}^2 + \frac{m^2}{1-z} + \frac{m_{\pi}^2}{z}} \,.
 \nn
\end{eqnarray} 
Thus, the Collins function is entirely given by the imaginary parts of 
the coefficients defined in Eqs.~(\ref{e:sig}--\ref{e:gamma2}).  
We can compute these imaginary parts by applying Cutkosky rules to the 
self-energy and vertex diagrams of Fig.~\ref{f:sigmagammadelta}. 
Explicit calculation leads to
\begin{eqnarray}
&& \im \big( A+B \big) = \frac{g_\sa^2}{32 \pi^2 F_\pi^2}
 \nn \\
&& \quad \mbox{} \times \bigg( 2m_\pi^2 - \frac{k^2-m^2}{2} 
 \Big(1 - \frac{m^2 - m_{\pi}^2}{k^2}\Big) \bigg) I_1 \,,
 \label{e:first} \\
&&\im \big( C_1 - mE_1 + (k^2-m^2)F_1 \big) = \frac{g_\sa^2}{32 \pi^2 F_\pi^2} 
 \nn \\
&& \quad \mbox{} \times  m \, (k^2-m^2) 
 \bigg( \frac{3k^2+m^2-m_\pi^2}{2 k^2} I_1 
 \\
&& \quad \mbox{}+ 4 m^2 \frac{k^2 -m^2 +m_{\pi}^2}{\lambda(k^2,m^2,m_{\pi}^2)} 
 \Big( I_1 + (k^2 - m^2 -2m_{\pi}^2) I_2 \Big) \bigg) \,,
 \nn \\
&&\im \big( C_2 - mE_2 + (k^2-m^2)F_2 \big) = \frac{1}{32 \pi^2 F_\pi^2} 
 \nn \\
&& \quad \mbox{} \times  m \, (k^2-m^2) 
 \bigg( 1 - \frac{m^2 - m_\pi^2}{k^2} \bigg) I_1 \,,
\end{eqnarray} 
where we have introduced the so-called K\"allen function, 
$\lambda(k^2,m^2,m_{\pi}^2)=[k^2 -(m+m_{\pi})^2][k^2 -(m-m_{\pi})^2]$,
and the factors
\begin{eqnarray}
I_1 & = & \int \de^4 l \; \delta (l^2 - m_{\pi}^2)\, \delta ((k-l)^2-m^2) \nn \\
 & = & \frac{\pi}{2 k^2} \sqrt{ \lambda(k^2,m^2,m_{\pi}^2)}
	\;\theta(k^2 -(m+m_{\pi})^2) \,, \\
I_2 & = & \int \de^4 l \; \frac{\delta(l^2 - m_{\pi}^2)\,\delta((k -l)^2-m^2)}
       {(k - p - l)^2 - m^2} \nn \\
 & = & -\frac{\pi}{2 \sqrt{\lambda(k^2,m^2,m_{\pi}^2)}}
	\ln{\left| 1+\frac{\lambda(k^2,m^2,m_{\pi}^2)}{k^2m^2 -(m^2-m_{\pi}^2)^2 }
	\right|} \nn \\ 
&&	\mbox{}\times 
	\theta(k^2 -(m+m_{\pi})^2) \,.
 \vphantom{\frac{1}{1}} 
 \label{e:last}
\end{eqnarray} 
These integrals are finite and vanish below the threshold of quark-pion 
production, where the self-energy and vertex diagrams do not possess an 
imaginary part.

Thus, Eq.~(\ref{e:h1}) in combination with 
Eqs.~(\ref{e:first})--(\ref{e:last}) gives the explicit result for the 
Collins function in the Manohar-Georgi model to lowest possible order. 
Because of its chiral-odd nature, the Collins function would vanish 
in this model if we set the mass of the quark to zero.
The same phenomenon has been observed in the calculation
of a chiral-odd twist-3 fragmentation function~\cite{Ji:1993qx}.
The result in Eq.~(\ref{e:h1}) corresponds, e.g., to the fragmentation 
$u \to \pi^0$. 
The expressions of the remaining favoured transitions are obtained in 
analogy to Eqs.~(\ref{e:flavour1},\ref{e:flavour2}).
Unfavoured fragmentation processes in the case of the Collins function 
appear only at the two-loop level.

\section{Estimates and phenomenology}
\label{s:three}

\subsection{Unpolarized fragmentation function and 
the choice of parameters}

We now present our numerical estimates, where all results for the fragmentation 
functions in this subsection refer to the transition $u \to \pi^+$.
To begin with we calculate the unpolarized fragmentation function $D_1(z)$ 
which is given by
\begin{equation} 
D_1(z)= \pi \int_0^{K^{2}_{\st\,{\rm max}}} \de K^2_{\st}\; D_1(z,K^2_{\st}),
\end{equation} 
where $\vec{K}_{\st}=- z \vec{k}_{\st}$ denotes the transverse momentum of the 
outgoing hadron with respect to the quark direction. 
The upper limit on the $K^2_{\st}$ integration is set by the cutoff on
the fragmenting quark virtuality, $\mu^2$, and corresponds to
\begin{equation} 
K^{2}_{\st\,{\rm max}}=z \, (1-z)\,\mu^2 -z\,m^2-(1-z)\,m_\pi^2 \,.
\end{equation} 
Besides $m$ and $g_\sa$, the cutoff $\mu^2$ is the third parameter of
our approach that is not fixed a priori.
However, as will be explained below, the possible values of $\mu^2$ can be
restricted when comparing our results to experimental data. 
Unless otherwise specified, we always use the values
\begin{equation} 
 m = 0.3 \; {\rm GeV}, \qquad g_{\sa}=1, \qquad \mu^2=1\; {\rm GeV}^2 \,.
 \label{e:param}
\end{equation} 
At the relevant places, the dependence of our results on possible variations 
of these parameters will be discussed.
Few remarks concerning the choice in (\ref{e:param}) are in order.
The value of $m$ is a typical mass of a constituent quark.
The choice for the axial coupling can be seen as a kind of average number
of what has been proposed in the literature.
For instance, in a simple SU(6) spin-flavour model for the proton one finds 
$g_\sa \approx 0.75$ in order to obtain the correct value for the axial charge 
of the nucleon~\cite{Manohar:1984md}.
On the other side, large $N_c$ arguments favour a value of the order 
of one~\cite{Weinberg:1990xn}, while, according to a recent calculation 
in a relativistic point-form approach~\cite{Boffi:2001zb}, a $g_\sa$ slightly
above one seems to be required for describing the axial charge of the nucleon.
Finally, our choice for $\mu^2$ ensures that the momenta of the outgoing pion
and quark, in the rest-frame of the fragmenting quark, remain below values
of the order $0.5 \, \rm {GeV}$.
In this region we believe chiral perturbation theory to be applicable, meaning
 that our leading order result should provide a reliable estimate.

\begin{figure}
        \centering
        \epsfig{figure=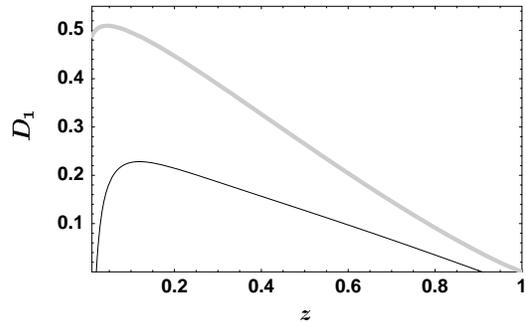,width=8cm}
        \caption{Model result for the unpolarized quark fragmentation 
	function $D_1^{u \rightarrow \pi^+}$ (solid line) and comparison with a 
	parametrization of Ref.~{\protect \cite{Kretzer:2000yf}} (grey line).}
	\label{f:d1}
        \end{figure}

In Fig.~\ref{f:d1} we show the result for the unpolarized fragmentation
function $D_1^{u \rightarrow \pi^+}$. 
Notice that in general the fragmentation functions vanish outside the 
kinematical limits, which in our model are given by
\begin{eqnarray} 
z_{{\rm max},{\rm min}}&=&
	\frac{1}{2}\Biggl[\left(1-\frac{m^2-m_{\pi}^2}{\mu^2}\right) \nn \\
	&&\qquad\pm \sqrt{\left(1-\frac{m^2-m_{\pi}^2}{\mu^2}\right)^2 -4\,
	\frac{m_{\pi}^2}{\mu^2}}\Biggr],
\end{eqnarray}   
corresponding to the situation when the upper limit of the $K_{\st}^2$ 
integration becomes equal to zero.
We consider our tree level result as a pure valence-type part of 
$D_1^{u \rightarrow \pi^+}$.
The sea-type (unfavoured) transition $\bar{u} \to \pi^+$ is strictly zero 
at leading order.
Therefore, we compare the model result to the valence-type quantity 
$D_1^{u\rightarrow \pi^+}-D_1^{\bar{u}\rightarrow \pi^+}$, where the
fragmentation functions have been taken from the parametrization of Kretzer
\cite{Kretzer:2000yf} at a scale $Q^2=1 \, \rm{GeV}^2$.\footnote{Other 
parametrizations~\protect{\cite{Kniehl:2000fe,Bourhis:2001gs,Kretzer:2001pz}} 
use
a starting energy scale $Q^2 \ge 2\, \rm{GeV}^2$, which is too high to allow a
comparison with our results.}
Obviously, the $z$-dependence of both curves is in nice agreement.
We point out that such an agreement is non-trivial.
For example, in the pseudoscalar model that we used in our previous 
work~\cite{Bacchetta:2001di}, $D_1$ behaves quite differently and peaks 
at an intermediate $z$-value. 

On the other side, we underestimate the parametrization of 
Ref.~\cite{Kretzer:2000yf} by about a factor of two. 
Some remarks are in order at this point.
Although a part of the discrepancy might be attributed to the uncertainty 
in the value of
$g_\sa$, the most important point is to address the question as to what extent
we can compare our estimate with existing parametrizations.
The parametrization of~\cite{Kretzer:2000yf} serves basically as input
function of the pQCD evolution equations, used to describe high-energy
$e^+ e^-$ data, and displays the typical logarithmic dependence on the scale
$Q^2$. A value of $Q^2 = 1$ GeV$^2$ is believed to be already beyond the
limit of applicability of pQCD calculations.
On the other side, our approach displays, to a first approximation, a linear
dependence on the cutoff $\mu^2$. It is supposed to be valid at low scales 
and it is also stretched to the limit of its applicability for
$\mu^2 = 1$ GeV$^2$.
In this context it should also be investigated to what extent the inclusion
of one-loop corrections, which allow for the additional decay channel
$q^{\ast} \to \pi \pi q$, will increase the result for $D_1$ at 
$\mu^2 = 1 \, \rm{GeV}^2$.
Finally, we want to remark that
to our knowledge there exists no strict one-to-one correspondence
between the quark virtuality $\mu^2$ and the scale used in the evolution 
equation of fragmentation functions, which in semi-inclusive DIS, e.g., 
is typically identified with the photon virtuality $Q^2$.
For all these reasons, 
a smooth matching of our calculation and the parametrization 
of~\cite{Kretzer:2000yf} cannot necessarily be expected.
Despite these caveats, the correct $z$-behaviour displayed by our result 
for $D_1$ suggests that 
the calculation can well be used as an
input for evolution equations at a low scale. 
In the next subsection we will elaborate more on this point in connection with
the Collins function. 

	\begin{figure}
        \centering
        \epsfig{figure=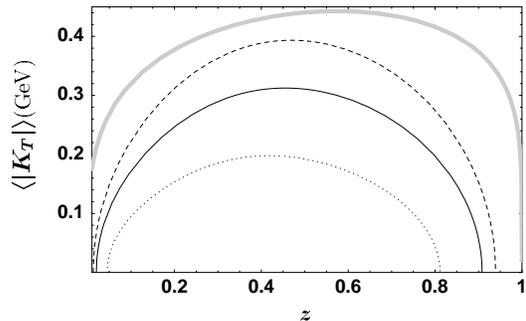,width=8cm}
        \caption{Model result for the average hadron transverse momentum 
	for different choices of the cutoff:  $\mu^2=0.5$ GeV$^2$ 
	(dotted line),
	$\mu^2=1$ GeV$^2$ (solid line), $\mu^2=1.5$ GeV$^2$ (dashed line)
	and comparison with a fit to experimental results from 
        DELPHI~{\protect \cite{Abreu:1996na}} (grey line).} 
	\label{f:trans}
        \end{figure}

The best indication of the appropriate value of the cutoff $\mu^2$ may be 
obtained when
comparing our calculation to experimental data of the average transverse 
momentum of the outgoing hadron with respect to the quark, which we evaluate
according to 
\begin{equation} 
\langle |\vec{K}_{\st}|\rangle (z)= \frac{\pi}{D_1(z)} 
	\int_0^{K^{2}_{\st\,{\rm max}}}
	\de K^2_{\st}\,|\vec{K}_{\st}|\, D_1(z, K^2_{\st}) \,.
\end{equation} 
In Fig.~\ref{f:trans} we show the result of this observable as a function of 
$z$ for three different choices of the parameter $\mu^2$. 
As a comparison, we also show a fit (taken from Ref.~\cite{Anselmino:1999pw}) to 
experimental data obtained by the DELPHI collaboration~\cite{Abreu:1996na}. 
Like in the case of $D_1(z)$, the shape of our result is very similar to the 
experimental one, which we consider as an encouraging result.
For $\mu^2 = 1\, \rm{GeV}^2$ our curve is about $30\%$ below the data.
Such a disagreement is not surprising, keeping in mind that at LEP energies
higher order pQCD effects (e.g. gluon bremsstrahlung, unfavoured
fragmentations, etc.) 
play an important role, leading in general to a broadening of the
$K_\st$ distribution.
For experiments at lower energies, however, where pQCD contributions 
can be neglected in a first approximation, it may be possible to exhaust 
the experimental value for $\langle |\vec{K}_{\st}|\rangle (z)$ 
with genuine soft 
contributions as described in our model.
This in turn would determine the appropriate value of the cutoff $\mu^2$.
For example, such a method of matching our calculation
with experimental conditions could be applied at HERMES kinematics, 
even though the method is somewhat
hampered since $K_\st$ is not directly measured in semi-inclusive DIS.
In this case, one rather observes the transverse momentum of the outgoing hadron 
with respect to the virtual photon, $P_{h \perp}$, which 
depends on both $K_\st$ and the transverse momentum of the 
partons inside the target $p_\st$.
At leading order in the hard scattering cross section one can in fact 
derive the relation
\begin{eqnarray} 
\lefteqn{\langle P^2_{h\perp}\rangle(x,z)} \nn \\
	&\qquad=& z^2 \frac{\pi \int
	\de p^2_{\st}\, p_{\st}^2\, f_1(x, p^2_{\st})}{f_1(x)}+\frac{\pi \int
	\de K^2_{\st}\, K_{\st}^2\, D_1(z, K^2_{\st})}{D_1(z)} \nn \\
	&\qquad=& z^2\,\langle p^2_{\st}\rangle (x) + \langle K^2_{\st}\rangle(z) \,,
\end{eqnarray} 
where $x$ represents the Bjorken variable.

\subsection{Collins function}

\begin{figure}
        \centering
        \epsfig{figure=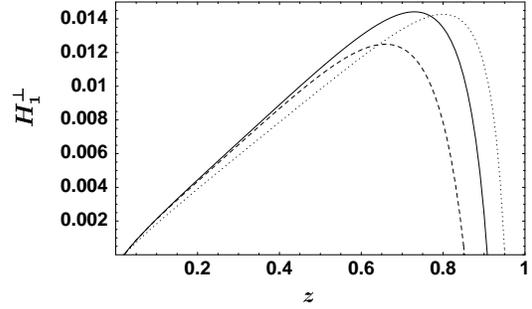,width=8cm}
        \caption{Model result for the Collins function for different values
	of the constituent quark mass: $m=0.2$ GeV (dotted line),
	$m=0.3$ GeV (solid line), $m=0.4$ GeV (dashed line).} 
	\label{f:collwithm}
        \end{figure}

We now turn to the description of our model result for the Collins function.
In Fig.~\ref{f:collwithm}, $H_1^\perp$ is plotted for three different values of
the constituent quark mass, $m= 0.2, \, 0.3, \, 0.4$ GeV. 
In a large $z$-range, the function does not depend strongly on the precise value 
of the quark mass, if we choose it within reasonable limits. 
That's why we can confidently fix $m = 0.3 \, \rm{GeV}$ for our numerical studies. 
It is very interesting to observe that the behaviour of the unpolarized
fragmentation function $D_1$ is quite distinct from that of 
the Collins function:
while the former is decreasing as $z$ increases, the latter is growing.

\begin{figure}
        \centering
        \epsfig{figure=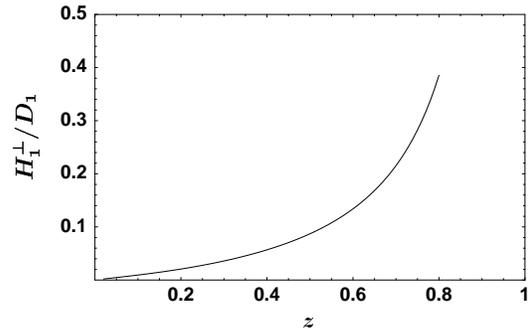,width=8cm}
        \caption{Model result for $H_1^{\perp}/D_1$.}
	\label{f:ratio}
        \end{figure}

The different behaviour of the two functions becomes even more evident
when looking at their ratio, shown in Fig.~\ref{f:ratio}.
We emphasize that also from the experimental side there exists some evidence
for an increasing ratio $H_1^\perp / D_1$.
In a recent analysis of the longitudinal single spin asymmetry measured at 
HERMES, Efremov {\it et al.}~\cite{Efremov:2001cz} extracted a behaviour
$H_1^\perp / D_1 \propto z$ for $z \le 0.7$.
We consider the agreement in finding a clearly rising ratio $H_1^\perp / D_1$ 
as remarkable, even though in the analysis of Ref.~\cite{Efremov:2001cz} 
some simplifying assumptions have been used in order to obtain information on 
$H_1^\perp$ from data taken with a longitudinally polarized target.
It will be very interesting to see if dedicated future experiments can confirm 
such a behaviour.
We also mention that ratios of the Collins function or any of its moments 
with $D_1$ are almost independent of the coupling constant $g_{\sa}$. 
The reason is that the one-loop correction containing the 
contact interaction is only proportional to $g_{\sa}^2$, as $D_1$ is, 
and is dominating on the 
others.
Furthermore, the ratio $H_1^{\perp}/D_1$ is nearly independent of the
cutoff $\mu^2$.
In conclusion, the prediction shown in Fig.~\ref{f:ratio} is almost
independent of the choice of parameters in our approach.

At this point we would like to add some general remarks concerning the
$z$-behaviour of our results.
It turns out that the shape of all the results does not vary much when
changing the parameters within reasonable limits.
In particular, variations of $g_\sa$ and of the cutoff $\mu^2$ only change
the normalization of the curves but not their shape.
In this sense our calculation of fragmentation functions has a strong
predictive power.
This has a direct practical consequence if one uses, for instance, our
result of the Collins function as input in an evolution equation:
the $z$-dependence of the input function can be adjusted to the shape of 
our $H_1^\perp$, while its normalization can be kept free in order to account
for uncertainties in the values of $g_\sa$ and $\mu^2$.

\begin{figure}
        \centering
        \epsfig{figure=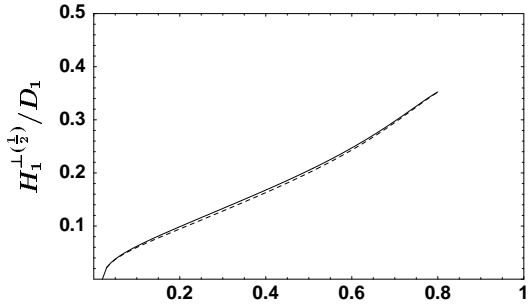,width=8cm}
        \caption{Model result for $H_1^{\perp (1/2)}/D_1$
	(solid line) and comparison with the product
	$(\langle |\vec{K}_{\st}|\rangle/2 z m_{\pi})\, (H_1^{\perp}/
	 D_1)$ (dashed line). 
	Note that the positivity bound requires the ratio to
	be smaller than $0.5$.} 
	\label{f:ratiomom0}
        \end{figure}

In Fig.~\ref{f:ratiomom0} we plot the ratio
\begin{equation}	
 \frac{H_1^{\perp
(\frac{1}{2})}(z)}{D_1(z)} \equiv
\frac{\pi}{D_1(z)} \int \de K_{\st}^2\, \frac{|\vec{K}_{\st}|}{2 z m_{\pi}}\,
H_1^{\perp}(z,K_{\st}^2) \,,
\label{e:ratiomom0}
\end{equation} 
which enters the transverse single spin asymmetry to be 
discussed in the following 
subsection.
This quantity rises roughly linearly within a large $z$-range, leading to a 
similar $z$-behaviour of the transverse spin asymmetry.
$H_1^{\perp(1/2)}/D_1$ is no longer independent of the cutoff $\mu^2$, but rather 
the same dependence as in the case of $\langle |\vec{K}_{\st}|\rangle$ 
(shown in Fig.~\ref{f:trans}) can be assumed.
In Fig.~\ref{f:ratiomom0}, this ratio  is compared to the expression
\begin{equation}	
 \frac{\langle |\vec{K}_{\st}|\rangle (z)}{2 z m_{\pi}}
	 \frac{H_1^{\perp}(z)}{D_1(z)} = 
\pi \,\frac{H_1^{\perp}(z)}{D_1^2(z)} \int \de K_{\st}^2 \frac{|\vec{K}_{\st}|}
{2 z m_{\pi}} 
D_1(z,K_{\st}^2) \,.
\end{equation}
A very close agreement between the two different curves can be observed,
indicating that the model predicts a
quite similar transverse momentum dependence of both the Collins function and $D_1$.
In the literature, this feature is sometimes assumed in phenomenological 
parametrizations of $H_1^\perp$.
Note, however, that in our approach deviations from this simple behaviour can be 
expected, if $D_1$ is also calculated consistently to the one-loop 
order.
	
The Collins function has to fulfill the positivity 
bound~\cite{Bacchetta:2000kz,Bacchetta:2000zm} 
\begin{equation}
\frac{|\vec{K}_{\st}|}{2 z m_{\pi}}\,
H_1^{\perp}(z,K_{\st}^2) \leq \frac{1}{2}\,D_1(z,K_{\st}^2).
\end{equation} 
Integration over $K_{\st}^2$ gives the simplified expression
\begin{equation}
\frac{H_1^{\perp (\frac{1}{2})}(z)}{D_1(z)}\leq \frac{1}{2} \, ,
\label{e:boundint}
\end{equation}	
which is satisfied by our model calculation. 
It is clear, however, that increasing the value of $\mu^2$ will eventually 
result in a violation of the positivity condition.
To avoid such a violation, we should calculate $D_1$ and $H_1^{\perp}$ 
consistently at the same order, i.e., the one-loop corrections to $D_1$ 
should be included.
By doing so, the positivity bound will be fulfilled even at larger values 
of $\mu^2$, for which our numerical results are no longer trustworthy.

\begin{figure}
        \centering
        \epsfig{figure=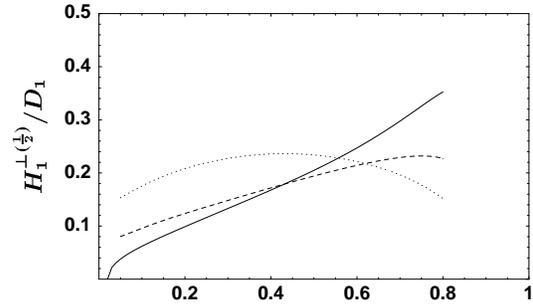,width=8cm}
        \caption{Model result for $H_1^{\perp (1/2)}/D_1$
	(solid line) and comparison with the same ratio, where 
        $H_1^{\perp(1/2)}$ is calculated according to 
        Eq.~(\ref{e:collansatz}) 
	with $M_C=0.3$ GeV (dashed line) and $M_C=0.7$ GeV (dotted line).}
	\label{f:collansatz}
        \end{figure}

From our results, we expect an increasing
behaviour of the azimuthal asymmetry in $p^\uparrow p\to\pi\,X$ as
function of $x_F$, qualitatively similar to what has been predicted 
in Ref.~\cite{Artru:1997bh} in the context of the Lund-fragmentation model. 
At this point, it is also interesting to discuss the comparison of our results 
with the ones obtained using the so-called ``Collins guess''. 
On the basis of very general assumptions, Collins suggested a possible
behaviour for the transverse spin asymmetry containing 
$H_1^\perp$~\cite{Collins:1993kk}. 
This suggestion has been used in the literature 
(see, e.g.,
Refs.~\cite{Oganessyan:1998ma,Kotzinian:1999dy,DeSanctis:2000fh,Ma:2001ie})  
to propose the following shape for the Collins function
\begin{equation}
 H_1^{\perp(\frac{1}{2})}(z) = \pi \int \de K_{\st}^2 \, 
\frac{|\vec{K}_\st|}{2 z} \,
\frac{M_C} {M_C^2 + \frac{K_{\st}^2}{z^2}}\, D_1(z,K_{\st}^2),
\label{e:collansatz}
\end{equation}	
with the parameter $M_C$ ranging between 0.3 and $0.7 \, \rm{GeV}$. 
Using our model outcome for the unpolarized fragmentation function, we apply
Eq.~(\ref{e:collansatz}) to estimate  $H_1^{\perp(1/2)}$, and in
Fig.~\ref{f:collansatz} we show how this compares to the exact result of
Eq.~(\ref{e:ratiomom0}).  
There is a rough qualitative agreement with the Collins ansatz for the lowest 
value of the parameter $M_C$, although it is not growing fast enough 
compared to the exact evaluation. 
On the other side, in the Manohar-Georgi model there is no agreement with 
the Collins ansatz for high values of the parameter $M_C$, which might indicate
that the relation suggested in Eq.~(\ref{e:collansatz}) should be 
handled with care.

\begin{figure}
        \centering
        \epsfig{figure=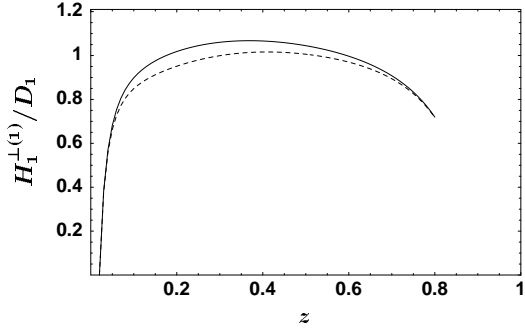,width=8cm}
        \caption{Model result $H_1^{\perp (1)}/D_1$
	(solid line) and comparison with the product
	$(\langle K_{\st}^2\rangle/2 z^2 m_{\pi}^2) (H_1^{\perp}/D_1^{\perp})$
	(dashed line).} 
	\label{f:ratiomom}
        \end{figure}

Finally, we display in Fig.~\ref{f:ratiomom} the quantity
\begin{equation}	
 \frac{H_1^{\perp(1)}(z)}{D_1(z)} \equiv
\frac{\pi}{D_1(z)} \int \de K_{\st}^2\, \frac{K_{\st}^2}{2 z^2 m_{\pi}^2}\,
H_1^{\perp}(z,K_{\st}^2) \,,
\end{equation} 
because this ratio appears in the weighted asymmetries to be considered below.
In Fig.~\ref{f:ratiomom}, also the expression
\begin{equation}	
 \frac{\langle K_{\st}^2\rangle (z)}{2 z^2 m_{\pi}^2}
	 \frac{H_1^{\perp}(z)}{D_1(z)} = 
\pi \,\frac{H_1^{\perp}(z)}{D_1^2(z)} \int \de K_{\st}^2 \frac{K_{\st}^2}
{2 z^2 m_{\pi}^2} 
D_1(z,K_{\st}^2) \,
\end{equation}
is shown for comparison.
Once again, there is a remarkable agreement between the two different
expressions, confirming the quite similar $K_\st$ behaviour of $H_1^\perp$
and $D_1$.

\subsection{Asymmetries in semi-inclusive DIS and $e^+e^-$ annihilation}

We turn now to estimates of possible observables containing the Collins
function. We will take into consideration one-particle inclusive DIS,
 where the Collins function appears in connection with the transversity
distribution of the nucleon, 
and  $e^+e^-$ annihilation into two hadrons belonging to two different jets.

In the first case, we consider the
DIS cross section with a transversely polarized target\footnote{
Transverse vectors and azimuthal angles are defined as lying on a plane 
perpendicular to the direction of the virtual photon.} 
and the production of one pion.
We denote the transverse polarization vector of the
target as $\vec{S}_T$. The cross
section is differential in six variables, for which we choose 
$x, y, z, |\vec{P}_{h\perp}|, \phi^S_h, \phi^S_l$, 
where $\vec{P}_{h\perp}$ is the transverse component of the
pion momentum, $\phi^S_h$ is
its azimuthal angle with respect to the target spin, and $\phi^S_l$ is the 
azimuthal angle of the lepton scattering plane again
with respect to the target spin. 

Orienting the spin of the target in two opposite directions and summing the
cross sections we isolate the unpolarized part~\cite{Mulders:1996dh}, 
\begin{eqnarray} 
&\lefteqn{\de^6\sigma_{U\uparrow} + \de^6\sigma_{U\downarrow}\;=\;
	\frac{4 \alpha^2_{\rm em}}{s x y^2} \,\left(1-y+\frac{y^2}{2}\right)}
		\nn \\
&\qquad&\qquad \mbox{}\times \int \de^2 \vec{p}_{\st} \de^2 \vec{k}_{\st} \; 
\delta^2 \left(\vec{p}_{\st}-\frac{\vec{P}_{h\perp}}{z} - \vec{k}_{\st}\right)
	 \nn \\
&&\qquad \mbox{}\times 
	\sum_a e^2_a f_1^a (x,p_{\st}^2)\, D_1^a (z,z^2 k_{\st}^2),
\label{e:sum}
\end{eqnarray} 
where the subscript $U$ indicates an unpolarized electron beam,
the index $a$ denotes quark flavours, and $f_1$ is the usual unpolarized
quark distribution in the nucleon. 
Subtracting the cross sections we obtain the polarized
part~\cite{Mulders:1996dh}, 
\begin{eqnarray} 
\lefteqn{\de^6\sigma_{U\uparrow} - \de^6\sigma_{U\downarrow}\;=\;
   - |\vec{S}_T|\,\frac{4 \alpha_{\rm em}^2}{s x y^2}\, (1-y)\, \sin{(\phi_h^S-2 \phi_l^S)}}
	 \nn \\
&\qquad& \mbox{} \times \int \de^2
\vec{p}_{\st} \de^2 \vec{k}_{\st} \;
\delta^2 \left(\vec{p}_{\st}-\frac{\vec{P}_{h\perp}}{z} - \vec{k}_{\st}\right)
	\nn \\
&& \mbox{} \times  \frac{\vec{P}_{h\perp} \cdot
\vec{k}_{\st}}{|\vec{P}_{h\perp}|\,m_{\pi}}\;\sum_a e^2_a h_1^a (x,p_{\st}^2)\,
H_1^{\perp\,a}(z,z^2 k_{\st}^2).
\label{e:diff}
\end{eqnarray} 
Integration over the azimuthal angles would cause the polarized part of the
cross section to vanish. After defining the angle $\phi \equiv
\phi_h^S-2 \phi_l^S$, we consider the $\sin \phi$ 
 weighted transverse spin asymmetry
\begin{eqnarray} 
\lefteqn{\left\langle \sin{\phi} \right\rangle_{UT} (x,y,z)} \nn \\
&\qquad=& 	\frac{\int  \de \phi_l^S\, \de^2 \vec{P}_{h\perp}\,
	\sin{\phi}\; (\de^6\sigma_{U\uparrow} - \de^6\sigma_{U\downarrow})}
	{\int  \de \phi_l^S\, \de^2 \vec{P}_{h\perp}
	(\de^6\sigma_{U\uparrow} + \de^6\sigma_{U\downarrow})}\, .
\end{eqnarray} 
Inserting Eqs.~(\ref{e:diff},\ref{e:sum}) into the
definition of the asymmetry results in an expression where the 
transverse momenta of $h_1$ and $H_1^{\perp}$ are still
entangled in a convolution integral~\cite{Ralston:1979ys}. 
To resolve the convolution, it is required to assume a
particular dependence of the transversity
distribution on the intrinsic transverse momentum. 
The simplest example is
\begin{equation}	
h_1 (x,p_{\st}^2) \approx h_1 (x) \, \frac{\delta (p_{\st}^2)}{\pi}\, ,
\end{equation} 
that means supposing there is no intrinsic transverse momentum of the partons 
inside the
target. Under this assumption, the pion transverse momentum with respect to
the virtual photon is entirely due
to the fragmentation process, i.e., $\vec{P}_{h\perp}=\vec{K}_T=-z \vec{k}_T$,
and the convolution can be disentangled 
\begin{eqnarray} 
\lefteqn{\left\langle \sin{\phi} \right\rangle_{UT} (x,y,z)}\nn \\
&\qquad \approx& |\vec{S}_{\st}|\frac{\frac{1}{x
y^2}\, (1-y)\,\sum_a e_a^2\, h_1^a (x)\,H_1^{\perp (\frac{1}{2})a}(z)}
{\frac{1}{x y^2}\left(1-y+\frac{y^2}{2}\right)\sum_a e_a^2\, f_1^a
(x)\,D_1^a(z)}\, ,
\label{e:asym1}
\end{eqnarray} 
where the approximation sign reminds that the equality is 
assumption-dependent. 

If we want to disentangle the convolution integral of Eq.~(\ref{e:diff}) 
without making any assumption on the intrinsic transverse 
momentum distribution, we need to weight the integral with the
magnitude of the pion transverse momentum~\cite{Kotzinian:1997wt}. 
This procedure results in 
the azimuthal transverse spin asymmetry
\begin{eqnarray} 
\lefteqn{\bigg\langle \frac{|\vec{P}_{h\perp}|}{m_{\pi}} \sin{\phi}
\bigg\rangle_{UT} (x,y,z) } \nn \\
&\qquad=& \frac{\int  \de \phi_l^S \de^2 \vec{P}_{h\perp}
\,\frac{|\vec{P}_{h\perp}|}{m_{\pi}}\,\sin{\phi}\;
(\de^6\sigma_{U\uparrow} - \de^6\sigma_{U\downarrow})}
{\int  \de \phi_l^S \de^2 \vec{P}_{h\perp}
(\de^6\sigma_{U\uparrow} + \de^6\sigma_{U\downarrow})} \nn \\
&\qquad=& |\vec{S}_{\st}|\frac{\frac{1}{x y^2}\,
	 (1-y)\,z\,\sum_a e_a^2\, h_1^a (x)\,H_1^{\perp (1)a}(z)}
{\frac{1}{x y^2}\,\left(1-y+\frac{y^2}{2}\right)\,\sum_a e_a^2\, f_1^a
(x)\,D_1^a(z)}\;. 
\label{e:asym2}
\end{eqnarray} 
We achieved an assumption-free factorization of the $x$ dependent transversity
distribution and the $z$-dependent Collins function. The measurement of this
asymmetry 
requires to bin the cross section according to the magnitude of the 
pion transverse momentum.
On the other side, this asymmetry represents potentially
the cleanest method to measure the transversity distribution together with
the Collins function. Moreover, it is 
not afflicted by
complications due to Sudakov factors~\cite{Boer:2001he}.

We show predictions for both
transverse spin asymmetries defined in
Eqs.~(\ref{e:asym1}) and (\ref{e:asym2}). Different calculations can be found
in the literature, e.g., 
in Refs.~\cite{Anselmino:2000mb,Korotkov:2001jx,Ma:2001ie}.
To estimate the magnitude of the asymmetries, 
we need inputs for the
distribution functions, in particular for the transversity distribution. 
Several model calculations of this function
are available at present (see \cite{Barone:2002sp} for a comprehensive
review). We refrain ourselves from considering many
different examples and rather restrict the analysis to two 
limiting situations. In the first case we adopt the
non-relativistic assumption $h_1 = g_1$, while in the second case we 
exhaust the upper bound on the transversity distribution, i.e., 
$h_1 \leq \frac{1}{2}(f_1 + g_1)$~\cite{Soffer:1995ww}. 
We use the simple parametrization
of $g_1$ and $f_1$ suggested in~\cite{Brodsky:1995kg}.
 At the moment, more sophisticated
parametrizations are available, taking also scale
evolution into account. 
However, all these parametrizations are compatible with each other to the 
extent of our purpose here, that is to give an estimate of the asymmetries
for a low scale. 
We focus on the production of $\pi^+$, where the contribution of
down quarks is negligible, not only because of the presence of unfavoured
fragmentation functions, but also because the transversity distribution 
for down
quarks is supposed to be much smaller than for up quarks.

\begin{figure}
        \centering
        \epsfig{figure=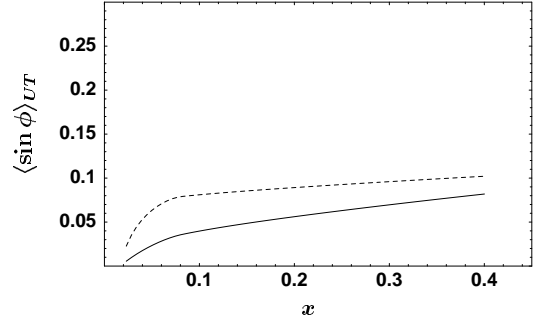,width=8cm}
        \caption{Azimuthal transverse spin asymmetry 
	$\langle\sin{\phi}\rangle_{UT}$ as a function of $x$.
	Solid line: assuming $h_1 = g_1$. Dashed line: assuming
	$h_1 =\frac{1}{2} (f_1+ g_1)$. The functions $f_1$ and
	$g_1$ are taken from {\protect\cite{Brodsky:1995kg}}.}
	\label{f:axnw}
        \end{figure}

\begin{figure}
        \centering
        \epsfig{figure=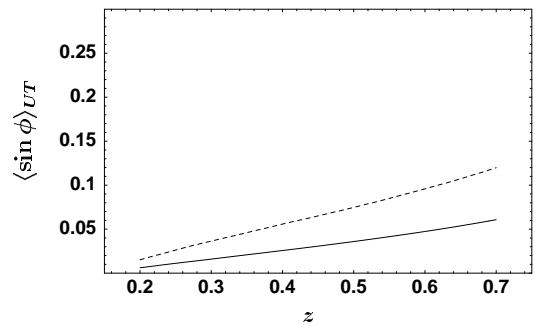,width=8cm}
        \caption{Azimuthal transverse spin asymmetry 
	$\langle\sin{\phi}\rangle_{UT}$ as a function of $z$.
	Solid line: assuming $h_1 = g_1$. Dashed line: assuming
	$h_1 =\frac{1}{2} (f_1+ g_1)$. The functions $f_1$ and
	$g_1$ are taken from {\protect\cite{Brodsky:1995kg}}.} 
	\label{f:aznw}
        \end{figure}

In Fig.~\ref{f:axnw} we present the azimuthal asymmetry
defined in Eq.~(\ref{e:asym1}) as a function of $x$, 
after integrating numerator and denominator over
the variables $y$ and $z$, for the two cases described above. 
In Fig.~\ref{f:aznw}, we present the same asymmetry as a function of $z$,
after integrating over $y$ and $x$. 
As already mentioned before, our prediction is supposed to be valid at a low
energy scale of about $1\, \rm{GeV}^2$. Neglecting evolution effects, 
it could be utilized for comparison with
experiments at a scale of few GeV$^2$.
We assume the value of the transverse
polarization to be $|\vec{S}_T| = 0.75$.
In performing the integrations, we apply
the kinematical cuts typical of the HERMES experiment, as described
in~\cite{Airapetian:2000tv}. Therefore, our prediction is particularly
significant for HERMES, which is supposed to be the first experiment 
to measure this asymmetry.
In principle, the simultaneous study of the $x$ and $z$-dependence of the
asymmetry yields separate information on the distribution and fragmentation
parts and allows to
extract both up to a normalization factor~\cite{Korotkov:2001jx}. Note,
however, that this procedure relies on the assumption of up-quark dominance 
and is valid only if the
asymmetry is truly factorized, so that the $x$-dependence can be ascribed
entirely to the distribution functions and the $z$-dependence entirely to the
fragmentation functions. Kinematical cuts could partially spoil this
situation.  
We would like to stress that our calculation predicts an asymmetry up to the
order of 10\%, which
should be within experimental reach, and suggests the possibility to
distinguish between different assumptions on the transversity distribution.

\begin{figure}
        \centering
        \epsfig{figure=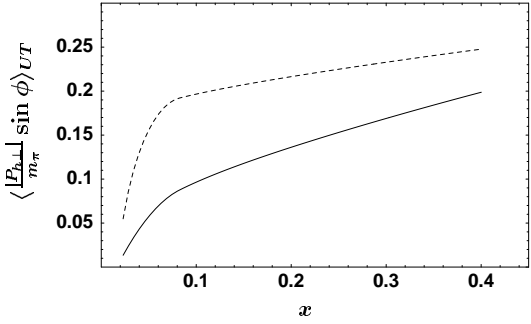,width=8cm}
        \caption{Azimuthal spin asymmetry 
	$\langle\frac{|\vec{P}_{h \perp}|}{m_{\pi}}\,\sin{\phi}\rangle_{UT}$ 
	as a function of $x$.
	Solid line: assuming $h_1 = g_1$. Dashed line: assuming
	$h_1 =\frac{1}{2} (f_1+ g_1)$. The functions $f_1$ and
	$g_1$ are taken from {\protect\cite{Brodsky:1995kg}}.}         
	\label{f:axw}
        \end{figure}

\begin{figure}
        \centering
        \epsfig{figure=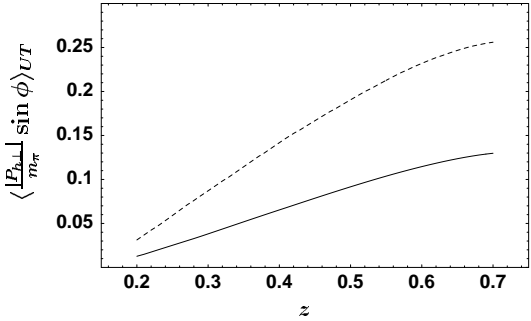,width=8cm}
        \caption{Azimuthal spin asymmetry 
	$\langle \frac{|\vec{P}_{h \perp}|}{m_{\pi}}\,\sin{\phi}\rangle_{UT}$
	as a function of $z$.
	Solid line: assuming $h_1 = g_1$. Dashed line: assuming
	$h_1 =\frac{1}{2} (f_1+ g_1)$. The functions $f_1$ and
	$g_1$ are taken from {\protect\cite{Brodsky:1995kg}}.} 
	\label{f:azw}
        \end{figure}

Using the same procedure as before, we have estimated the asymmetry defined in
Eq.~(\ref{e:asym2}), containing the weighting with 
$|\vec{P}_{h \perp}|/m_{\pi}$. 
The results are shown in Fig.~\ref{f:axw} as a function of $x$  and  
in Fig.~\ref{f:azw} as a function of $z$. The magnitude of this asymmetry is
higher than in the unweighted case, which is due to the
fact that the weighting spuriously enhances the asymmetry by about a factor 
two.  

Besides appearing in semi-inclusive DIS in connection with the transversity
distribution of the nucleon, the Collins function can be independently 
extracted from another
process, that is
electron-positron 
annihilation into two hadrons belonging to two
back-to-back jets~\cite{Boer:1997mf,Boer:1998qn}. We restrict ourselves to the
case of $\gamma$-exchange only.
In this process, one of the two hadrons
(say hadron 2) defines the scattering plane together with the leptons and 
determines the direction with respect to which the azimuthal 
angles must be
measured. The cross section is differential in five variables, e.g.,
$z_1, z_2, y, |\vec{P}_{h \perp}|, \phi$. The variables $z_1$ and
$z_2$ are the longitudinal fractional momenta of the two hadrons. In the
center of mass frame $y=(1+\cos{\theta})/2$, where $\theta$ is the angle of
hadron 2 with respect to the momentum of the incoming leptons. The vector
$\vec{P}_{h \perp}$ denotes the transverse component of the momentum of hadron
1 and $\phi$ is its azimutal angle with respect to the scattering plane. For a
more detailed description of the kinematical variables we refer to 
\cite{Boer:1997mf,Boer:1998qn}.

We define the azimuthal asymmetry  
\begin{eqnarray} 
\lefteqn{\left\langle P_{h \perp}^2 \cos{2 \phi}
\right\rangle_{e^+ e^-} (\theta,z_1,z_2) } \nn \\
&=& \frac{\int  \de^2 \vec{P}_{h\perp}
\,P_{h\perp}^2\,\cos{2 \phi}\;
\de^5\sigma_{e^+ e^-}}
{\int \de^2 \vec{P}_{h\perp} \,P_{h\perp}^2\,
\de^5\sigma_{e^+ e^-}}  \\
&=& \frac{2 \, \sin^2 {\theta}}{1+\cos^2{\theta}}\,
\frac{H_1^{\perp (1)} (z_1)\,\bar{H}_1^{\perp (1)}(z_2)}
{\left( D_1 (z_1)\,\bar{D}_1^{(1)}(z_2)
	+ D_1^{(1)} (z_1)\,\bar{D}_1(z_2)\right)},  \nn
\end{eqnarray} 
where
summations over quark flavours are understood.	
The weighting with a second power of $P_{h \perp}$ in the numerator is 
necessary to obtain a deconvoluted expression. We prefer to use the same
weighting in the denominator as well, to avoid a modification of the 
asymmetry just caused by the weighting. 

\begin{figure}
        \centering
        \epsfig{figure=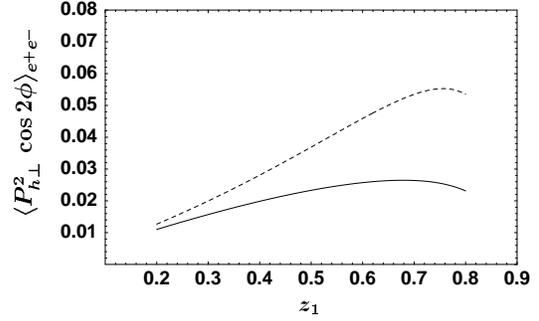,width=8cm}
       \caption{Azimuthal asymmetry 
	$\langle P_{h \perp}^2 \cos{2 \phi}\rangle_{e^+ e^-}$ 
	for $e^+ e^-$ annihilation into two hadrons, 
	integrated over the range $0.2 \leq z_2 \leq 0.8$ (solid line), and
	over the range $0.5 \leq z_2 \leq 0.8$ (dashed line).} 
	\label{f:aee}
        \end{figure}

In Fig.~\ref{f:aee} we present the estimate of the asymmetry defined above,
entirely based on our model. The asymmetry has been integrated over $z_2$ and
$\theta$, leaving the dependence on $z_1$ alone. 
We have extended the $\theta$ integration interval all the way to $[0,\pi]$, to
obtain a conservative estimate. In fact, limiting the interval to
$[\frac{\pi}{4},\frac{3 \pi}{4}]$ will enhance the asymmetry by a factor 2,
approximately.
Because the Collins function increases with
increasing $z$, we also get a larger asymmetry by restricting
the integration range for $z_2$. 
As an illustration of this feature, in Fig.~\ref{f:aee} we present
two results, obtained from two different integration ranges.
Our prediction is supposed to
be valid only at low energy scales and should be evolved for comparison with
higher energy experiments. It is important to note that we estimate the 
asymmetry to be of the order of about 5\%, 
and thus should be very well observable in experiments.

\section{Summary and conclusions}
\label{s:four}

We have estimated the Collins fragmentation function for pions at a low
energy
scale by means of the Manohar-Georgi model.
This model 
contains three essential features of non-perturbative QCD:
massive quark degrees of freedom, chiral symmetry and 
its spontaneous breaking (with pions as Goldstone bosons). 
Because of the chiral invariant interaction between pions
and
quarks, the fragmentation process can be evaluated in a perturbative expansion.
The constituent quark
mass,
the axial pion-quark coupling $g_\sa$ and the maximum virtuality $\mu^2$
of the
fragmenting quark are free parameters of our approach.
The quark mass and $g_\sa$ 
are constrained within reasonable limits.
To ensure the convergence of the chiral perturbation expansion, $\mu^2$
cannot exceed a typical hadronic scale.
We have mostly considered the value $\mu^2 = 1\,\rm{GeV}^2$, which
guarantees that
the momenta of the particles produced in the fragmentation process
stay well below the scale of chiral symmetry breaking, $\Lambda_\chi \approx 1$
GeV.
To determine the appropriate value of $\mu^2$, 
the average transverse momentum of a data set could
be used.
In any case, we observed that variations of the free parameters within 
reasonable limits have only a minor
influence on the shape of the results, implying that
our approach has a strong 
predictive power for
the $z$-behaviour of the various functions.

We have found that the Manohar-Georgi model reproduces reasonably well the
unpolarized
pion fragmentation function and the average transverse momentum of a
produced hadron
as function of $z$, supporting the idea of describing the fragmentation
process
by a chiral invariant approach.

Compared to the unpolarized fragmentation function, modeling the
Collins function
is considerably more difficult, mainly because of its chiral-odd and
time-reversal
odd nature.
In our approach, the helicity-flip required to generate a chiral-odd
object
is caused by the mass of the constituent quark, while
the T-odd behaviour is produced via one-loop corrections.
The Collins function exhibits a quite distinct behaviour from the
the unpolarized fragmentation
function.
In particular, the ratio $H_1^\perp / D_1$ is strongly increasing with
increasing $z$.

On the basis of our results, we have calculated
the transverse single-spin asymmetry in semi-inclusive DIS 
where the Collins function appears in combination with the
transversity
of the nucleon.
This observable will be measured in the near future at HERMES and could
also be
investigated at COMPASS, Jlab (upgraded) and eRHIC.
For typical HERMES kinematics the asymmetry is of the order of $10\%$, giving
support to the
intention of extracting the nucleon transversity in this way.
We believe that our estimate of the Collins
function, despite its uncertainties,
can be very useful for this extraction.

More information on the Collins function from the experimental side is
urgently
required.
In this respect, the most promising experiment seems to be $e^+ e^-$
annihilation into two hadrons, where
$H_1^\perp$ appears squared in an azimuthal $\cos{2\phi}$ asymmetry.
According to our calculation, an asymmetry of the order of $5\%$ can be
expected, which should be measurable at high luminosity
accelerators, such as BABAR and BELLE.
Dedicated measurements of the Collins function would be extremely
important for
the extraction of the transversity distribution.
Moreover, they could answer the question whether a chiral invariant
Lagrangian can be
used to model the Collins function.

\acknowledgements

Discussions with D.~Boer and K.~Oganessyan are gratefully acknowledged. 
This work is part of the research program of the Foundation for Fundamental 
Research on Matter (FOM) and the Netherlands Organization for Scientific 
Research (NWO), and it is partially funded by the European Commission IHP 
program under contract HPRN-CT-2000-00130.


\end{document}